\documentclass[10pt,journal,letterpaper]{IEEEtran}

\usepackage{nopageno}
\usepackage{geometry}
\usepackage{graphicx} 
\usepackage{cite}
\usepackage{amsthm,amsmath,amssymb,amsfonts}
\usepackage{enumerate} 
\usepackage{caption}

\newtheorem{lemma}{Lemma}
\newtheorem{theorem}{Theorem}
\newtheorem{remark}{Remark}

\DeclareMathOperator{\rank}{rank}
\DeclareMathOperator{\sinr}{SINR}
\DeclareMathOperator{\prob}{Prob}
\DeclareMathOperator{\diag}{diag}
\DeclareMathOperator{\tr}{tr}
\DeclareMathOperator{\minimize}{min.}

\DeclareMathOperator{\subjectto}{s.t.}
\DeclareMathOperator{\argmin}{argmin}

\DeclareMathOperator{\E}{\mathsf{E}}
\DeclareMathOperator{\Cov}{\mathsf{cov}}

\begin{document}
\thispagestyle{empty}
\def\BibTeX{{\rm B\kern-.05em{\sc i\kern-.025em b}\kern-.08em
 M   T\kern-.1667em\lower.7ex\hbox{E}\kern-.125emX}}
\title{Goal-Oriented Communication for Networked Control Assisted by Reconfigurable Meta-Surfaces}
\author{Mohamad Assaad and Touraj Soleymani
\thanks{Mohamad Assaad is with the Laboratory of Signals and Systems, CentraleSup\'{e}lec, University of Paris-Saclay, 91190 Gif-sur-Yvette, France. Touraj Soleymani is with the City St George's School of Science and Technology, University of London, London EC1V~0HB, United Kingdom.}%
}
\maketitle

\thispagestyle{empty}
\begin{abstract}
In this paper, we develop a theoretical framework for goal-oriented communication assisted by reconfigurable meta-surfaces in the context of networked control systems. The relation to goal-oriented communication stems from the fact that optimization of the phase shifts of the meta-surfaces is guided by the performance of networked control systems tasks. To that end, we consider a networked control system in which a set of sensors observe the states of a set of physical processes, and communicate this information over an unreliable wireless channel assisted by a reconfigurable intelligent surface with multiple reflecting elements to a set of controllers that correct the behaviors of the physical processes based on the received information. Our objective is to find the optimal control policy for the controllers and the optimal phase policy for the reconfigurable intelligent surface that jointly minimize a regulation cost function associated with the networked control system. We characterize these policies, and also propose an approximate solution based on a semi-definite relaxation technique.
 \end{abstract}

\begin{IEEEkeywords}
Control actions, networked control systems, optimal policies, phase shifts, reconfigurable intelligent surfaces.
\end{IEEEkeywords}

\section{Introduction}
Reconfigurable meta-surfaces, a.k.a.~reconfigurable intelligent surfaces (RISs), are artificial planar structures with integrated electronic circuits, which are equipped with large numbers of scattering elements capable of modifying wireless propagation environments~\cite{liu2021rec}. Our conviction is that integration of the RIS technology into the control technology holds immense potential to revolutionize the landscape of connected robotic systems. In these networked control systems, where real-time exchange of information between sensors, controllers, and actuators is essential for effective regulation of physical processes, the adoption of reconfigurable meta-surfaces can significantly enhance the quality of communication. The RISs achieve this enhancement by intelligently redirecting and focusing signals, reducing blockage, mitigating interference, and providing dynamic adaptation and resource allocation, thereby ensuring prompt delivery of data packets to their intended destinations.

In this paper, we develop a theoretical framework for goal-oriented communication assisted by reconfigurable meta-surfaces in the context of networked control systems. The relation to goal-oriented communication~\cite{uysal2022semantic, voi, voi2, touraj-thesis, soleymani2023state, soleymani2024found, soleymani2024consis, maatouk2022age, kriouile2021global} stems from the fact that optimization of the phase shifts of the meta-surfaces is guided by the performance of networked control systems tasks. To that end, we consider a networked control system in which a set of sensors observe the states of a set of physical processes, and communicate this information over an unreliable wireless channel assisted by a RIS with multiple reflecting elements to a set of controllers that correct the behaviors of the physical processes based on the received information. This study is motivated by a multitude of practical applications, e.g., smart factories~\cite{bhatia2024exp}, internet-of-thing devices~\cite{niu2022joint}, autonomous vehicles~\cite{ozcan2021reconf}, and unmanned~aerial~vehicles~\cite{yang2022perf}.

On one hand, there are previous studies that have explored the performance of RIS-assisted systems over wireless channels in terms of coverage, signal to noise ratio (SNR) and ergodic capacity. Notably, a single-user wireless system assisted by a RIS was considered in~\cite{ruizhang19}, where a joint active and passive beamforming was develop to minimize the total transmit power. In addition, a multi-user wireless system assisted by a RIS was considered in~\cite{zyang21}, where a joint transmit base-station power and passive beamforming were developed. For a survey on this topic, we refer the reader to~\cite{poor23}. On the other hand, there are previous studies that have investigated the performance and the stability of networked control systems over wireless channels. Notably, in the seminal work in~\cite{sinopoli}, mean-square stability of Kalman filtering over an independent and identically distributed (i.i.d.) erasure channel was studied, where it was proved that there exists a critical point for the packet error rate above which the expected estimation error covariance is unbounded. Moreover, mean-square stability of Kalman filtering over a fading channel with correlated gains was investigated in~\cite{quevedo2013}, where a sufficient condition that ensures the exponential boundedness of the expected estimation error covariance was established. For a survey on this topic, we refer the reader to \cite{park2017wireless}. Nevertheless, we should emphasize that, to the best of our knowledge, no prior research has been conducted on potential interactions between RISs, sensors, and controllers in networked control~systems. 

This paper introduces a novel framework for the integration of the RIS technology into the control technology. We mathematically model a networked control system consisting of multiple sensors, multiple controllers, and a RIS that assist the communication between the sensors and the controllers. Our main objective is to find the optimal control policy for the controllers and the optimal phase policy for the RIS that jointly minimize a regulation cost function associated with the networked control~system. The paper is organized as follows. The mathematical model of our networked system and the problem formulation are presented in Section II. We derive the optimal policies and propose an approximate solution in Section III. We provide the numerical results in Section IV, and conclude the paper in Section V.

\section{Problem Statement}
We consider a RIS-assisted networked control system consisting of $K$ sensors, $K$ controllers, and a RIS with $M$ reflecting elements that assist the communication between the sensors and the controllers. The time $t$ is discrete, and the time horizon $T$ is finite. In our setup, the $k$th controller is responsible for regulating the behavior of the $k$th dynamical process, whose evolution is described~as
\begin{align}
	{x}_{k}(t+1)={A}_k {x}_k(t) +{B}_k {u}_k(t) +{w}_k(t)\label{state_evolution}
\end{align} 
for $k \in [1,K]$ and $t \in [0,T]$, where ${x}_k(t) \in \mathbb{R}^n$ is the state of the $k$th dynamical process at time $t$, ${u}_k(t) \in \mathbb{R}^m$ is the control action of the $k$th controller, and ${w}_k(t)$ is a zero mean i.i.d Gaussian noise with covariance ${W}_k>0$, ${A}_k \in \mathbb{R}^{n \times n}$ and ${B}_k \in \mathbb{R}^{n \times m}$ are the state and input matrices of the $k$th dynamical process. We assume that ${x}_{k}(0)$ is a Gaussian vector with zero mean and covariance $P_k(0)$. At each time $t$, the $k$th sensor measures a noisy output of the $k$th dynamical process, which is modeled~as
\begin{equation}
	{y}_{k}(t)={C}_k {x}_k(t) + {v}_k(t)
\end{equation} 
for $k \in [1,K]$ and $t \in [0,T]$, where ${y}_{k}(t) \in \mathbb{R}^p$ is the output of the $k$th dynamical process at time~$t$, ${v}_k(t)$ is a zero mean i.i.d Gaussian noise with covariance ${V}_k>0$, and ${C}_k \in \mathbb{R}^{p\times p}$ is the output matrix of the $k$th dynamical process. 

At each time $t$, the $k$th sensor transmits its MMSE state estimate, represented by $\hat{{x}}^s_k(t)$, to the $k$th controller. This message is transmitted as a signal over the wireless channel that is subject to fading. At the $k$th controller, we consider that $\hat{{x}}^s_k(t)$ is decoded correctly with probability $1-P_e^k$, and otherwise it is dropped with probability $P_e^k$. The transmitted signal is reflected by the RIS according to the phase shifts ${\phi}(t)= [\phi_1(t), \dots,\phi_M(t)]^T$ such that $\phi_i(t) \in [0,2\pi]$. The received signal, therefore, depends on~${\phi}(t)$. 

\begin{remark}
We assume that the RIS's phase shifts depend on the statistical knowledge of the dynamical processes, i.e., ${\phi}(t)$ is independent of the realizations of ${x}_{k}(s)$ for $k \in [1,K]$ and $s \in [0,t]$. This assumption dramatically simplifies the deployment of the RIS, as it eliminates the need for persistent access to the sensor outputs. We also assume that, upon a successful delivery of a message by the $k$th controller at each time, an acknowledgment is sent to the $k$th sensor and the RIS via ideal feedback channels.
\end{remark}

The channel between the $k$th sensor and the $k$th controller at time $t$ is in fact a concatenation of following components: the direct link between the $l$th sensor and $k$th controller $h^{sc}_{lk}(t) \in \mathbb{C}$, the link between the $k$th sensor to the RIS $h^{sr}_k(t) \in \mathbb{C}^M$, and the link between the RIS and the $k$th controller $h^{rc}_k(t) \in \mathbb{C}^M$, for $k \in \{1,\dots, K\}$. Therefore, the received signal at the $k$th controller is given by 
\begin{equation}
{r}_{k}(t) = \sum_{l=1}^K\left(h^{sc}_{lk}(t)+h^{sr}_l(t) {\Theta}(t) h^{rc}_k(t)\right) {s}_l(t)+{n}_k(t)
\end{equation} 
for $k \in [1,K]$ and $t \in [0,T]$, where ${r}_{k}(t)$ is the signal received by the $k$th controller at time~$t$, ${s}_{k}(t)$ is the signal transmitted by the $l$th sensor, ${\Theta}(t)=\diag(e^{j\phi_1(t)},\dots,e^{j\phi_M(t)}) \in \mathbb{C}^{M \times M}$ is the the RIS's phase shift matrix, and ${n}_k(t)$ is the zero mean i.i.d. Gaussian noise with covariance ${N}_k>0$. Observe that, at the $k$th controller, the useful signal power is $S_k(t) = p_k \| h^{sc}_{kk}(t) + h^{sr}_k(t) {\Theta}(t) h^{rc}_k(t) \|^2$ and the interference power is $I_k(t)=  \sum_{l=1,l\neq k}^K p_l \| h^{sc}_{lk}(t) + h^{sr}_l(t) {\Theta}(t) h^{rc}_k(t) \|^2$, where $p_l$ is the power of the transmitted signal ${s}_l(t)$. 

\begin{remark}
We assume that the RIS's phase shifts depend only on the statistical CSI information, i.e., $\phi(t)$ is independent of the realizations of $h^{sc}_{lk}(s)$, $h^{sr}_k(s)$, and $h^{rc}_k(s)$ for $l,k \in [1,K]$ and $s \in [0,t]$. This assumption leads to a significant reduction in the complexity and signaling overhead in the RIS deployment. We also assume that the transmission powers of all sensors are fixed and normalized, i.e., $p_k(t) = 1$ for $k\in[1,K]$ and $t\in[0,T]$.
\end{remark}

In this networked system, the decision variables are phase shifts and control actions. We represent a phase policy~$\Phi$ by $\{ \phi_1(t),\dots ,\phi_M(t) \}_{t=0}^{T}$ and a control policy $\Psi$ by $\{ {u}_1(t), \dots, {u}_K(t) \}_{t=0}^{T}$. Our main objective is to find a phase policy and a control policy that jointly minimize the cost function, i.e., we would like to solve
\begin{align}
&\underset{\Phi, \Psi}{\minimize} \ \E \bigg[\sum_{k=1}^K \!\Big(\!\sum_{t=0}^{T} {x}^T_k(t) {D}_k {x}_k(t)+ \sum_{t=0}^{T-1} {u}^T_k(t) {E}_k {u}_k(t) \!\Big) \bigg] \label{cost2}  \\
&\subjectto \ \ \phi_i(t) \in [0, 2\pi] \ \ \forall i=1,..,M \label{constraintphi}
\end{align}
where ${D}_k \in \mathbb{R}^{n \times n}$ and ${E}_k \in \mathbb{R}^{m \times m}$ are two predefined positive definite matrices associated with the $k$th dynamical process.

\begin{remark}
The cost function in the above optimization problem is the linear-quadratic-regulator cost function, which is widely used for measuring the performance of control systems. In our problem, this cost function should be minimized collaboratively by the controllers and the RIS. Note that the weighting matrices ${D}_k$ and ${E}_k$ specify the relative criticality of the dynamical processes with respect to one another, and also the relative importance of the state deviations with respect to the control efforts in each dynamical process.
\end{remark}

\section{Optimal Solutions}
In this section, we first derive the optimal control actions by applying the separation principle of control theory. Building on this control policy, we present a reduced optimization problem in terms of the phase shifts. We derive the optimal phase shifts, and propose an approximate solution based on one-step lookahead policy, an upperbound for the packet error rate, and a semi-definite relaxation.

\subsection{Optimal Control Actions}
Note that the information available at the $k$th sensor at time~$t$ can be described by the information set
\begin{align}
\mathcal{I}_k^s(t) = \Big\{{y}_k(0), \dots,{y}_k(t),{u}_k(0), \dots,{u}_k(t\!-\!1) \Big\} .
\end{align}
Accordingly, we can define the following state estimates and error covariances at the $k$th sensor at time $t$: $\hat{{x}}^s_k(t|t) = \E[{x}_k(t)|\mathcal{I}_k^s(t)]$, $\hat{{x}}^s_k(t|t\!-\!1) = \E[{x}_k(t)|\mathcal{I}_k^s(t\!-\!1)]$, $\hat{{P}}^s_k(t|t) = \Cov[{x}_k(t)|\mathcal{I}_k^s(t)]$, and $\hat{{P}}^s_k(t|t\!-\!1) = \Cov[{x}_k(t)|\mathcal{I}_k^s(t\!-\!1)]$.

At each time $t$, the $k$th sensor transmits $\hat{{x}}^s_k(t|t)$ through the RIS-mediated channel to the $k$th controller. Let $\delta_k(t)$ be an indicator variable such that $\delta_k(t)=1$ if the information transmitted by the $k$th sensor is decoded correctly by the $k$th controller, and $\delta_k(t)=0$ otherwise. Note that $\delta_k(t)$ depends only on the phase shift ${\phi}(t)$ such that $\mathbb{P}\left(\delta_k(t)=0|{\phi}(t)\right) = P^e_k({\phi}(t))$ and $\mathbb{P}\left(\delta_k(0)=0, \dots, \delta_k(t)=0|{\phi}(0), \dots,{\phi}(t)\right)=\mathbb{P}\left(\delta_k(0)=0|{\phi}(0)\right) \dots \mathbb{P}\left(\delta_k(t)=0|{\phi}(t)\right)$.   
In addition, the information available at the controller at time $t$ can be described by the information set 
\begin{align}
\mathcal{I}_k^c(t) = \Big\{ {z}^s_k(0|0), \dots,{z}^s_k(t|t),{u}_k(0), \dots,{u}_k(t\!-\!1) \Big\}
\end{align}
where $z_k^s(t|t) = \delta_k(t)\hat{{x}}^s_k(t|t)$. Accordingly, we can define the following state estimates and error covariances at the $k$th controller at time $t$: $\hat{{x}}^c_k(t|t) = \E[{x}_k(t)|\mathcal{I}_k^c(t)]$, $\hat{{x}}^c_k(t|t\!-\!1) = \E[{x}_k(t)|\mathcal{I}_k^c(t\!-\!1)]$, $\hat{{P}}^c_k(t|t) = \Cov[{x}_k(t)|\mathcal{I}_k^c(t)]$, and $\hat{{P}}^c_k(t|t\!-\!1) = \Cov[{x}_k(t)|\mathcal{I}_k^c(t\!-\!1)]$.

\begin{lemma}
The optimal control policy $\Psi^\star$ is given by
\begin{align}
{u}^\star_k(t)={L}_k(t) \hat{{x}}^c_k(t|t)    
\end{align}
where
\begin{align}
&\hat{{x}}^c_k(t|t) = \delta_k(t) \hat{{x}}^s_k(t|t)+\big( 1-\delta_k(t) \big) {A}_k \hat{{x}}^c_k(t\!-\!1|t\!-\!1) \label{estimate_controller} \\[1\jot]
&{L}_k(t) =- \big( {B}_k^T {\Omega}_k(t\!+\!1) {B}_k + {E}_k \big)^{-1} {B}_k^T {\Omega}_k(t\!+\!1) {A}_k \nonumber
\end{align}
and ${\Omega}_k(t)$ is a solution of the equation
\begin{align}
&{\Omega}_k(t) = {A}_k^T {\Omega}_k(t\!+\!1) {A}_k + {D}_k - {A}_k^T {\Omega}_k(t\!+\!1) {B}_k \nonumber\\[1\jot]
&\qquad \ \times \big({B}_k^T {\Omega}_k(t\!+\!1) {B}_k + {E}_k \big)^{-1}{B}_k^T {\Omega}_k(t\!+\!1) {A}_k . \label{ARE}
\end{align}
\end{lemma}
\begin{IEEEproof}
From (\ref{state_evolution}), we observe that $\hat{x}_k^c(t|t)$ can be written, depending whether the $k$th controller has correctly decoded the transmitted message at time $t$ or not, as
\begin{align*}
\hat{x}_k^c(t|t)= \begin{cases}
{A}_k \hat{x}_k^c(t\!-\!1|t\!-\!1) & \delta_k(t)=0 \\[1\jot]
\hat{x}_k^s(t|t) & \delta_k(t)=1
\end{cases}
\end{align*}
which is equivalent to (\ref{estimate_controller}). 

Now, following similar arguments as in \cite{leong2017event}, $\delta_k(s)$ are independent of the $x_k(s')$ for $k \in [1,K]$ and $s,s' \in [0,t]$, we can show that the design of the control policy and the phase policy becomes separated. Moreover, following similar arguments as in \cite{stoccontrol}, since $u_k(s)$ do not affect $\hat{P}^c_k(s')$ for $k \in [1,K]$ and $s,s' \in [0,t]$, we can show that separation principle of the control theory holds, and the control action at each time $t$ is obtained as ${u}^\star_k(t)= {L}_k(t) \hat{{x}}^c_k(t|t)$, where $L_k(t)$ is the linear-quadratic-regulator gain and (\ref{ARE}) is the associated algebraic Riccati equation.
\end{IEEEproof}

\begin{remark}
The optimal policy characterized by Lemma~1 is of a certainty-equivalent form. According to this policy, each controller only needs to substitute its MMSE state estimate at each time in the structure of the corresponding optimal state-feedback control policy.
\end{remark}

\subsection{Reduced Optimization Problem}
By adopting the results of Lemma 1, we can find a reduced optimization problem in terms of phase shifts, which is introduced in the next lemma.
\begin{lemma}
Let the control actions be given according to Lemma 1. The optimal phase shifts must satisfy the following optimization problem:
\begin{align}
&\underset{\Phi}{\minimize} \ \sum_{t=0}^{T-1} \sum_{k=1}^K \tr \big( {F}_k(t) \bar{{P}}^c_k(t|t) \big)\\[2\jot]
&\subjectto \ \ \phi_i(t) \in [0, 2\pi] \ \ \forall i=1,..,M\\[2\jot]
&\qquad \qquad \ \ \ \bar{{P}}^c_k(t|t) = f\big(\bar{{P}}_k^c(t\!-\!1|t\!-\!1), P^e_k({\phi}(t))\big)
\end{align}
where $\bar{{P}}_k^c(t|t) = \E[{P}_k^c(t|t) ]$ and
\begin{align}
&{F}_k(t) = {A}_k^T{\Omega}_k(t\!+\!1) {A}_k + {D}_k - {\Omega}_k(t) \label{F_matrix}\\[2\jot]
&f \big(\bar{{P}}_k^c(t\!-\!1|t\!-\!1), P^e_k({\phi}(t)) \big) = \hat{P}_k^s(t|t) \nonumber\\[2\jot]
&+\Big({A}_k \bar{{P}}_k^c(t\!-\!1|t\!-\!1) {A}_k^T + {W}_k - \hat{P}_k^s(t|t) \Big) P^e_k({\phi}(t)) \label{f_function}\\[1\jot]
&\hat{P}_k^s(t|t) = \Big( \big( A_{k}(t\!-\!1) \hat{P}_k^s(t\!-\!1|t\!-\!1) A_k^T(t\!-\!1) \nonumber\\[0\jot]
&\qquad \qquad + W_{k}(t\!-\!1) \big)^{-1} + C_k(t)^T V_k(t)^{-1} C_k(t) \Big)^{-1} .  \label{cov_KF}
\end{align}
\end{lemma}
\begin{IEEEproof}
Analogous to the derivation in \cite{schenato}, by plugging ${u}^\star_k(t)$ into the cost function in \ref{cost2}, we can write the cost function~as
\begin{align}
&\sum_{k=0}^{K} \tr \big( {\Omega}_k(0) \hat{P}^c_k(0) \big) + \sum_{t'=0}^{s} \sum_{k=0}^{K}\tr \big( {\Omega}_k(t'\!+\!1) {W}_k \big) \nonumber\\[-1\jot]
&+ \sum_{t'=0}^{s}\sum_{k=1}^K \tr \big( {F}_k(t') \E [\hat{P}_k^c(t'|t') ] \big) \label{cost-to-come}
\end{align}
for $s = T-1$, where the first two terms are independent of the phase shifts, and can be discarded.

From (\ref{state_evolution}), we observe that $\hat{P}_k^c(t|t)$ can be written, depending whether the $k$th controller has correctly decoded the transmitted message at time $t$ or not, as
\begin{align*}
\hat{P}_k^c(t|t)= \begin{cases}
{A}_k \hat{P}_k^c(t\!-\!1|t\!-\!1){A}_k^T + {W}_k & \delta_k(t)=0 \\[1\jot]
\hat{P}_k^s(t|t) & \delta_k(t)=1 .
\end{cases}
\end{align*}
Using this equation, we find
\begin{align*}
\bar{{P}}^c_k(t|t) &= \hat{P}_k^s(t|t) \Big(1-P^e_k({\phi}(t))\Big)\\[0\jot]
&+\Big({A} \bar{{P}}_k^c(t\!-\!1|t\!-\!1)]{A}^T +{W} \Big) P^e_k({\phi}(t))
\end{align*}
which is equivalent to (\ref{f_function}). Moreover, (\ref{cov_KF}) is the error covariance equation of the Kalman filter.
\end{IEEEproof}

\begin{remark}
The optimization problem in Lemma~2 depends only on the dynamics of the expected values of error covariances at the controllers, and should be optimized only with respect to the phase policy. 
\end{remark}

\vspace{-4mm}
\subsection{Optimal Phase Shifts}
Note that the information available at the RIS at time~$t$ can be described by the information set
\begin{align}
\mathcal{I}^{r}(t) \!=\! \Big\{\delta_k(0), \dots,\delta_k(t\!-\!1), {\phi}_k(0), \dots,{\phi}_k(t\!-\!1) \Big\} .
\end{align}
Given this information, the RIS can compute $\bar{{P}}^c_k(t\!-\!1|t\!-\!1)$ at time $t$ by applying (\ref{f_function}) and (\ref{cov_KF}). Define a value function associated with the reduced optimization problem from the perspective of the RIS as
\begin{align}
&V \big(t, \{ \bar{{P}}^c_k(t\!-\!1|t\!-\!1) \}_{k=1}^{K} \big) = \min_{{\phi}(t), \dots, {\phi}(T)} \nonumber\\[-1\jot]
&\qquad \sum_{t'=t}^{T} \sum_{k=1}^K \tr\Big( {F}(t')  f \big(\bar{{P}}_k^c(t'\!-\!1|t'\!-\!1), P^e_k({\phi}(t')) \big) \Big) .
\end{align}

The following theorem provides a general procedure for obtaining the optimal phase shifts.
\begin{theorem}
The optimal phase policy $\Phi^\star$ is given by
\begin{align}
&\phi^\star(t) = \underset{{\phi}(t)}{\argmin} \bigg\{\sum_{k=1}^K \tr\Big( {F}(t) f \big(\bar{{P}}_k^c(t\!-\!1|t\!-\!1), P^e_k({\phi}(t)) \big)\! \Big) \nonumber\\[-1\jot]
&\ \ \ + V \Big(t\!+\!1, \big\{ f \big(\bar{{P}}_k^c(t\!-\!1|t\!-\!1), P^e_k({\phi}(t)) \big) \big\}_{k=1}^{K} \Big) \bigg\}.
\end{align}
\end{theorem}

\begin{remark}
Finding the optimal phase shifts according to Theorem 1 can be quite challenging, due to the following reasons:  (i) the relation $P^e_k({\phi}(t))$ is often analytically unknown, and (ii) the value function $V \big(t, \{ \bar{{P}}^c_k(t\!-\!1|t\!-\!1) \}_{k=1}^{K} \big)$ should be solved backward in time. 
\end{remark}

\vspace{-4mm}
\subsection{Suboptimal Phase Shifts}
The signal to interference and noise ratio (SINR) at the $k$th controller is written as
$\sinr_k(t)= S_k(t)/(I_k(t)+N_0)$, where $N_0$ is the noise power. Let $R_k$ be the bit rate at which the signal is transmitted by the $k$th sensor. Then, $P^k_e({\phi}(t))$ can be written given as $P^k_e({\phi}(t))=\prob(\log_2(1+\sinr_k(t))<R_k)$, which is equivalent to $\prob(\sinr_k(t)<2^{R_k}-1)$. This can be justified by the fact that the maximum bit rate at which a signal can be transmitted without error is asymptotically given by the channel capacity. Therefore, an error occurs when the bit rate is higher than the capacity.  Consequently, 
\begin{align*}
&P^k_e({\phi}(t)) = \prob \big(S_k(t) \!-\!(2^{R_k} \!- \! 1)I_k(t)<N_0(2^{R_k} \!-\! 1) \big) \nonumber\\[1\jot]
&=\prob \big(\Gamma_k \!- \!S_k(t)+(2^{R_k} \!- \! 1)I_K(t) >\Gamma_k \!-\!N_0(2^{R_k} \!- \! 1) \big)
\end{align*}
where $\Gamma_k$ is the maximum value of $S_k(t)-(2^{R_k}-1)I_k(t)$. Since $S_k(t)-(2^{R_k}-1)I_k(t)>0$, and by using the Markov inequality, we get 
\begin{align}
P^k_e({\phi}(t) \hspace{-0.5mm}) & =\! \prob \hspace{-0.5mm} \big(\Gamma_k\! \hspace{-0.5mm} -\!S_k(t)\!+\!(2^{R_k}\hspace{-1.5mm}-\!1)I_k(t) \!> \! \Gamma_k \hspace{-0.5mm} \!-\!N_0(2^{R_k}\hspace{-1.5mm}-\!1) \hspace{-0.5mm} \big) \nonumber\\[1\jot]
& \leq \frac{\Gamma_k-\E[S_K(t)-(2^{R_k}-1)I_k(t)]}{\Gamma_k-N_0(2^{R_k}-1)}. \label{markov}
\end{align}
Note that, according to one-step lookahead policy, for the RIS we need to solve at time $t$:
\begin{align}
&\underset{\phi(t)}{\minimize} \ \sum_{k=1}^K \tr \big({F}_k(t) {G}_k(t) \big)  P^k_e({\phi}(t))
\end{align}
where ${G}_k(t) = {A}_k \bar{{P}}_k^c(t\!-\!1|t\!-\!1) {A}_k^T + {W}_k - \hat{P}_k^s(t|t)$. By adopting (\ref{markov}) in the above optimization problem, we get
\begin{align*}
&\underset{\phi(t)}{\minimize} \ \sum_{k=1}^K \tr \big({F}_k(t) {G}_k(t) \big) \frac{\Gamma_k \!-\!\E[S_K(t) \!-\!(2^{R_k}\!-\!1)I_k(t)]}{\Gamma_k\!-\!N_0(2^{R_k}\!-\!1)} .
\end{align*}

Now, we simplify the expectation term. Define ${H}_{kk}(t)=\diag\left((h^{sr}_k(t))^H\right)h^{rc}_k(t)$, ${H}_{lk}(t)=\diag\left((h^{sr}_l(t))^H\right)h^{rc}_k(t)$, and ${\theta}(t)=[\theta_1(t), \dots,\theta_M(t)]^T$. Then,
\begin{align*}
\E\left[S_k(t)\right] &= {\theta}(t)^H \E\left[{H}_{kk}(t){H}^H_{kk}(t)\right]{\theta}(t)+\E\left[\|h^{sc}_{kk}(t)\|^2\right]  \nonumber\\[2\jot]
&\quad +{\theta}(t)^H \E\left[{H}_{kk}(t) h^{sc}_{kk}(t)\right] + \E\left[h^{sc}_{kk}(t){H}^H_{kk}(t)\right]{\theta}\\[-1\jot]
\E\left[I_k(t)\right] &= \hspace{-2mm}\sum_{l=1,l\neq k}^K \hspace{-2mm} {\theta}(t)^H \E\left[{H}_{lk}(t){H}^H_{lk}(t)\right]{\theta}(t) + \E\left[\|h^{sc}_{lk}(t)\|^2\right]\nonumber\\[-1\jot]
&\quad + {\theta}^H(t) \E\left[{H}_{lk}(t) h^{sc}_{lk}(t)\right] + \E\left[h^{sc}_{lk}(t){H}^H_{lk}(t)\right]{\theta}(t).
\end{align*}

Note that $h^{sc}_{kk}(t)$ is the direct link between the $k$th sensor and the $k$th controller; hence, it is independent of ${H}_{kk}(t)$. Now, define $Q^{(1)}_k=\E[{H}_{kk}(t){H}^H_{kk}(t)]-(2^R_k-1)\sum_{l=1,l\neq k}^K\E[{H}_{lk}(t){H}^H_{lk}(t)]$, $Q^{(2)}_k=\E[{H}_{kk}(t) h^{sc}_{kk}(t)]-(2^R_k-1)\sum_{l=1,l\neq k}^K\E[{H}_{lk}(t) h^{sc}_{lk}(t)]$, $Q^{(3)}_k=\E[h^{sc}_{kk}(t){H}^H_{kk}(t)]-(2^R_k-1)\sum_{l=1,l\neq k}^K\E[h^{sc}_{lk}(t){H}^H_{lk}(t)]$, and $\Delta_k=\E[|h^{sc}_kk(t)\|^2]-(2^R_k-1)\sum_{l=1,l\neq k}^K\E[|h^{sc}_{lk}(t)\|^2]$. Define also the matrices $Q_k=[Q^{(1)}_k, Q^{(2)}_k; Q^{(3)}_k, 0]$,  $\tilde{\theta}(t)= [{\theta}(t); 1]$ and ${\Sigma}(t)=\tilde{\theta}(t)\tilde{\theta}(t)^H$.

By using that $\tilde{\theta}(t)^H Q_k \tilde{\theta}(t)=\tr\left(Q_k {\Sigma}(t)\right)$,  the phase shift optimization problem at time $t$ can then be written as
 \begin{align}
&\underset{\phi(t)}{\minimize} \ \sum_{k=1}^K  \tr \big({F}_k(t) {G}_k(t) \big) \frac{\Gamma_k-\Delta_k - \tr\left(Q_k {\Sigma}(t)\right)}{\Gamma_k-N_0(2^{R_k}-1)} \\[2\jot]
&\subjectto \ \ {\Sigma}(t) \succeq 0 ; \ \ \ \rank\left({\Sigma}\right)=1;\\[1\jot]
&\qquad \  \ {\Sigma}_{ii}(t)=1 \ \ \forall i=1, \dots,M . 
\end{align}

The above optimization problem can be solved using a semi-definite relaxation, which consists in dropping first the rank constraint and solving the following standard semi-definite programming (SDP):
 \begin{align}
&\underset{\phi(t)}{\minimize} \ \sum_{k=1}^K \tr \big({F}_k(t) {G}_k(t) \big) \frac{\Gamma_k-\Delta_k - \tr\left(Q_k {\Sigma}(t)\right)}{\Gamma_k-N_0(2^{R_k}-1)}\\[2\jot]
&\subjectto \ \ {\Sigma}(t) \succeq 0; \ {\Sigma}_{ii}(t)=1 \ \ \forall i=1, \dots,M . 
\end{align}

Since the solution cannot ensure that $\rank\left({\Sigma}(t)\right)=1$, a Gaussian randomization technique can be used to obtain a rank one solution~\cite{ruizhang19}.
This is done in the following way. Let ${\Sigma}^\star(t)$ be the optimal solution of the above SDP problem at time $t$. We generate a zero-mean complex Gaussian vector $g$ of size $M+1$ with covariance ${\Sigma}^\star(t)$, and select the first $M$ elements of this vector. Then, the phase shift solution is given~by ${\theta}_i^\star(t)=(g)_i/ \| (g)_i\|$ for all $i=1, \dots,M $, where $(g)_i$ is the $i$th element of the complex vector $g$ and $\| (g )_i\|$ is the norm of this $i$th element. The ratio $(g)_i/ \| (g)_i\|$ ensures that $\|{\theta}_i^\star(t)\|=1$.

\section{Numerical Results}
For our numerical analysis, we concentrated on a networked control system composed of a single RIS and two sensor-controller pairs. The RIS is placed at the origin of a two-dimensional Euclidean space. The sensors are place in the first quadrant randomly at distances ranging from $5 \text{m}$ to $20 \text{m}$. Moreover, the controllers are placed in the second quadrant randomly at distances ranging from $30 \text{m}$ to $70 \text{m}$. We used a Rician channel model, and assumed that the dynamical processes are scalar. The state coefficient $A_k$ is chosen randomly between $0.5$ and $10$ while the other parameters $B_k$, $C_k$, $P_k(0)$, $W_k$, $V_k$, $D_k$, and $E_k$ are set to $1$. The time horizon $T$ is set to $30$ slots. We compared the performance of the proposed approximate phase policy with that of the random phase policy in Fig.~1. In this diagram, the total cost-to-come at each time $t$ is defined according to (\ref{cost-to-come}) when $s = t$. We can observe that our method clearly achieves better performance.

\begin{figure}
	\centering
	\includegraphics[width=.94\columnwidth, trim={12mm 4mm 10mm 6mm},clip]{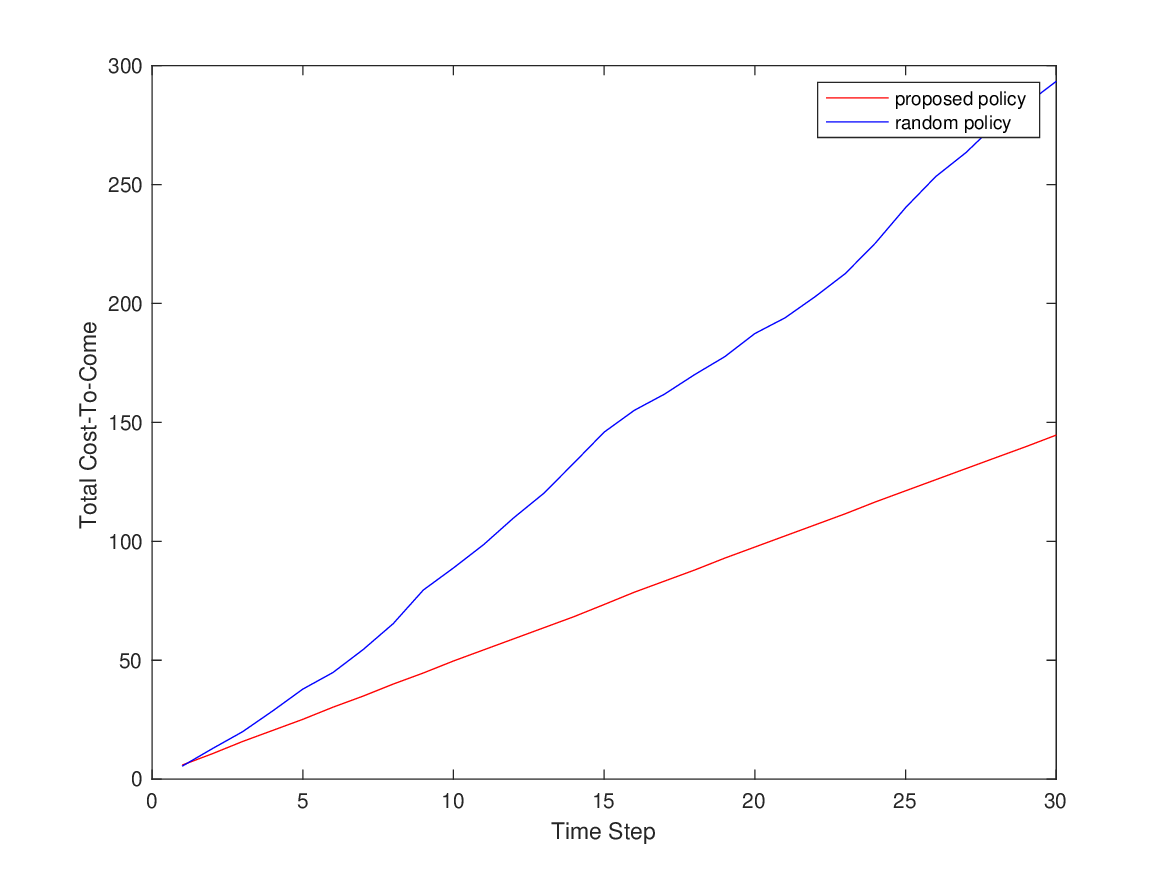}
	\caption{The time evolution of the total cost-to-come.}
	\label{fig:system}
\end{figure}

\section{Conclusion}
In this paper, we investigated the integration of a RIS in a networked control system. To analyze the interactions between communication and control, we developed a theoretical framework that enables us to find the optimal control policy and the optimal phase policy, such that they jointly minimize a regulation cost function. We characterized these optimal policies, and proposed a suboptimal SDP-based solution that can be readily implemented. Our numerical results confirmed the superiority of our solution compared to a~benchmark.

\bibliography{RIS.bib}
\bibliographystyle{ieeetr}

\end{document}